\documentclass[12pt,a4paper]{article}
\usepackage[utf8]{inputenc}
\usepackage{amsmath,amsfonts,amssymb}
\usepackage{graphicx}
\usepackage{authblk}
\usepackage[top=2.5cm,bottom=2.5cm,left=2.5cm,right=2.5cm]{geometry}
\usepackage{setspace}
\usepackage{cite}
\title{\textbf{Physics Opportunity at the Electron-Ion Collider (EIC)}}
\author[1]{Satoshi Yano}
\affil[1]{Hiroshima University, Japan}
\date{}

\begin{document}

\maketitle

\begin{abstract}
The Electron-Ion Collider (EIC) is a next-generation facility under construction at Brookhaven National Laboratory, uniquely designed to collide polarized electrons with polarized protons and ions. With its high luminosity, broad kinematic reach, and precise control over beam polarizations, the EIC will provide transformative insights into Quantum Chromodynamics (QCD). It will enable three-dimensional imaging of partons inside nucleons, reveal the role of gluons in the nucleon’s mass and spin, and probe gluon saturation at small Bjorken-$x$. Furthermore, the EIC will perform detailed measurements of gluon distributions in nuclei and explore how QCD dynamics evolve in nuclear environments. The ePIC detector, currently under development, is tailored to fully exploit these capabilities through advanced tracking, calorimetry, and particle identification across wide rapidity ranges. The EIC’s unique complementarity with heavy-ion programs at RHIC and LHC makes it a central tool for completing the QCD picture of visible matter.
\end{abstract}

\section{The Electron-Ion Collider Facility}
The Electron-Ion Collider (EIC) construction at Brookhaven National Laboratory is envisioned as the world’s first collider capable of accelerating and colliding polarized electron beams with either polarized protons or a variety of ion species. It will operate over a wide center-of-mass energy range, from 20 to 141~GeV, covering both low- and high-$x$ regimes essential to unraveling the structure of nucleons and nuclei~\cite{Accardi:2012qut}.

A distinguishing feature of the EIC is its high degree of flexibility: both electron and hadron beams can be configured in polarized or unpolarized modes, and nuclear beams can span a broad spectrum of mass numbers, from light nuclei like deuterons ($A = 2$) up to heavy ions such as uranium ($A = 238$). In terms of performance, the machine is designed to deliver instantaneous luminosities between $10^{33}$ and $10^{34}$~cm$^{-2}$s$^{-1}$, nearly three orders of magnitude higher than the HERA $e + p$ collider~\cite{HERA}, enabling extremely high-precision studies.

These advanced capabilities will empower the EIC to address a broad array of fundamental questions in quantum chromodynamics (QCD). Specifically, it will facilitate precision mapping of the three-dimensional structure of both nucleons and nuclei~\cite{Accardi:2012qut}, offer new insight into how the features of nucleons and nuclei emerge from their fundamental constituents, quarks and gluons.


\section{Scientific Objectives}
The EIC aims to explore several fundamental questions: the detailed origins of nucleon spin and mass, the gluonic structure of nucleons and nuclei, and the search for new states of matter predicted by QCD under extreme conditions~\cite{Aschenauer:2017jsk}. These topics are critical for a deeper comprehension of QCD and its implications for the structure of visible matter in the universe.

\subsection{Three-dimensional Structure of Nucleons}
A key scientific goal of the EIC is a complete mapping of the internal structure of the nucleon, including spatial and momentum distributions of its constituent partons. The ultimate description of such structure is given by the Wigner distribution, which encodes the simultaneous dependence on transverse spatial position ($\vec{b}_T$) and transverse momentum ($\vec{k}_T$) of partons. However, due to the uncertainty principle, the Wigner distribution cannot be directly observed.

Instead, this information is accessed through two complementary theoretical frameworks: Generalized Parton Distributions (GPDs)~\cite{Ji:1996nm} and Transverse Momentum Dependent distributions (TMDs)~\cite{Collins:2011zzd}. GPDs describe the correlation between the longitudinal momentum and transverse position of partons and are accessed in exclusive processes such as deeply virtual Compton scattering (DVCS) and deeply virtual meson production (DVMP). TMDs, on the other hand, characterize the transverse momentum distribution of partons and their spin correlations, and are typically probed in semi-inclusive deep inelastic scattering (SIDIS).

The EIC, by providing highly polarized electron beams and a broad range of kinematics, will enable precision measurements of spin asymmetries and azimuthal modulations in these processes. These observables are sensitive to the subtle interplay between parton transverse motion and nucleon spin, allowing experimental access to quantities such as the Sivers~\cite{Sivers:1990fh} and Boer--Mulders~\cite{Boer:1997nt} functions. Moreover, by combining GPD and TMD information, the EIC will contribute to a unified three-dimensional imaging of the nucleon in phase space, laying the foundation for tomographic studies of hadron structure.

\subsection{Mass Decomposition of the Nucleon}
A fundamental open question in QCD concerns the origin of the nucleon mass. Although the nucleon has a mass of approximately 938~MeV, the contribution from the Higgs-generated current quark masses is surprisingly small---only around 9\%. The remainder must arise from nonperturbative QCD dynamics, including kinetic and potential energy of quarks and gluons, as well as the quantum anomaly associated with scale symmetry breaking, known as the trace anomaly~\cite{Ji:1995sv}.

The mass decomposition can be analyzed in terms of the expectation value of the QCD energy-momentum tensor $T^{\mu\nu}$ in the nucleon state~\cite{Ji:1994av}. Of particular interest is the trace $T^{\mu}_{\ \mu}$, which, in the chiral limit, is dominated by the gluonic term arising from the renormalization of the QCD coupling.

Experimentally accessing this gluonic structure is highly challenging. However, the EIC will enable such measurements through the exclusive production of heavy quarkonium states, such as $\mathrm{J/}\psi$ and $\Upsilon$, in the near-threshold region~\cite{Hatta:2018ina}. These processes are sensitive to the gluon gravitational form factors, which are Fourier transforms of matrix elements of $T^{\mu\nu}$ and encode information about the pressure, shear forces, and energy distributions inside the nucleon.

The EIC's unprecedented luminosity and wide kinematic coverage will allow detailed studies of these processes, thereby providing direct constraints on the gluonic contributions to mass generation in QCD. Combined with lattice QCD results and phenomenological modeling, EIC data will significantly advance our understanding of how mass emerges from the underlying fields of the strong interaction.

\subsection{Nuclear Structure and Gluon Distributions}
The behavior of gluons inside nuclei remains one of the least understood aspects of QCD, especially in the regime of small Bjorken-$x$. In high-energy nuclear collisions, understanding the nuclear parton distribution functions (nPDFs) is essential for interpreting phenomena such as jet quenching and quark-gluon plasma formation. However, current constraints on nuclear gluon distributions, particularly at small-$x$, are limited due to the lack of precise experimental data.

The EIC will uniquely enable high-statistics measurements of DIS in various nuclear targets across a wide kinematic range. Through inclusive, semi-inclusive, and exclusive measurements, it will map out the gluon content in nuclei with unprecedented precision. Inclusive DIS provides access to the nuclear structure function $F_2^A$ and its scaling violations, which are sensitive to gluon distributions via DGLAP evolution~\cite{Eskola:2009uj}. Semi-inclusive measurements with identified final-state hadrons or jets offer sensitivity to the flavor and momentum distributions of partons within nuclei. Exclusive processes such as coherent vector meson production (e.g., $\mathrm{J/}\psi$ photoproduction) will probe the spatial distribution of gluons in the transverse plane~\cite{Mantysaari:2016ykx}.

One of the central goals of these measurements is to search for gluon saturation, a non-linear regime of QCD where gluon recombination balances gluon splitting at small-$x$, resulting in a saturation of gluon densities~\cite{Gelis:2010nm}. This regime is theoretically described by the Color Glass Condensate (CGC), an effective field theory that captures the coherent dynamics of dense gluonic matter. The EIC's ability to vary the nuclear size and collision energy will provide powerful leverage in testing the onset of saturation and exploring the universal properties of the CGC.

Moreover, measurements of diffractive cross sections in $e + A$ collisions, which are expected to constitute a significant fraction of the total cross section at small-$x$, will serve as sensitive probes of the coherence and spatial fluctuations of gluon fields. By comparing diffractive to inclusive event rates, the EIC will enable direct determination of the nuclear diffractive structure function and elucidate the role of color coherence effects in nuclei.

Altogether, the EIC's comprehensive nuclear program will transform our understanding of gluon distributions in nuclei, provide insights into emergent QCD phenomena, and offer essential inputs for interpreting heavy-ion collision results at RHIC and LHC.

\section{Experimental Setup}
A central element of the EIC scientific program is the ePIC detector, optimized for full event reconstruction across a wide kinematic range in $e+p$ and $e+A$ collisions~\cite{AbdulKhalek:2021gbh}. One of the unique features of DIS at the EIC is the strong boost toward the ion-going direction (forward rapidity) due to the asymmetric beam energies. Consequently, different regions of rapidity correspond to distinct physics processes and detector requirements.

The scattered electron, which defines the momentum transfer $Q^2$ and Bjorken-$x$, is typically observed in the electron-going (backward) direction, covering pseudorapidities $\eta < 0$ (in the lab frame). This region demands high-resolution electromagnetic calorimetry and tracking for precise lepton identification and momentum measurement. The electron identification is achieved via a combination of tracking detectors and finely segmented electromagnetic calorimeters, which allow for excellent energy resolution and $e/\pi$ discrimination.

In contrast, final-state hadrons produced in semi-inclusive and exclusive processes populate a wide rapidity range, extending from the central region ($|\eta| \lesssim 1$) to the forward ion-going direction ($\eta > 1$). To address this, the ePIC detector incorporates particle identification (PID) technologies tailored to different momentum and angular regimes. These include a barrel DIRC system (hpDIRC) for PID in the central region, a dual-radiator Ring Imaging Cherenkov (dRICH) detector in the forward direction for separating $\pi$, $K$, and $p$ at higher momenta, and a high-resolution Time-of-Flight (TOF) system using AC-LGADs for lower-momentum PID.

For neutral particle and jet energy measurements, the calorimeter system includes both electromagnetic calorimeters (EMCal) and hadronic calorimeters (HCal) with hermetic coverage. The EMCal in the electron-going direction is designed for high granularity to ensure accurate electron energy measurements and $\gamma$/$\pi^0$ separation. In the forward region, the calorimeters must handle high-energy jets and photon showers with fine timing and spatial resolution.

Altogether, the ePIC detector is designed to provide full acceptance in rapidity and azimuth, ensuring efficient detection of both scattered leptons and produced hadrons. This comprehensive coverage is essential for fulfilling the EIC's goals of unraveling QCD dynamics at the partonic level in both nucleons and nuclei.

\section{Conclusion}
The EIC is expected to serve not only as a precision microscope for the partonic structure of nucleons and nuclei, but also as a complementary tool to heavy-ion programs at RHIC and the LHC. While heavy-ion collisions probe QCD matter at high temperature and density—such as the quark-gluon plasma—the EIC offers the ability to study the initial-state structure and dynamics under controlled conditions. This synergy is particularly valuable for understanding the global picture of QCD phenomena. Its complementarity to existing and future heavy-ion facilities will make it a central pillar in the exploration of the strong interaction.

\end{document}